\documentclass[prl,twocolumn,floats,aps,epsfig,nofootinbib,axodraw,amssymb]{revtex4}
\usepackage{amsmath,amsfonts}
\usepackage{subfigure}
\usepackage{graphicx}
\usepackage{color}
\def\beq{\begin{equation}}
\def\eeq{\end{equation}}
\def\bea{\begin{eqnarray}}
\def\eea{\end{eqnarray}}

\def\leqn#1{(\ref{#1})}

\def\pslash{\not{\hbox{\kern-4pt $p$}}}
\def\qslash{\not{\hbox{\kern-4pt $q$}}}
\def\lv{\not{\hbox{\kern-4pt $L$}}}
\def\lsim{\mathrel{\raise.3ex\hbox{$<$\kern-.75em\lower1ex\hbox{$\sim$}}}}
\def\gsim{\mathrel{\raise.3ex\hbox{$>$\kern-.75em\lower1ex\hbox{$\sim$}}}}
\def\ifmath#1{\relax\ifmmode #1\else $#1$\fi}

\usepackage{graphicx}
\usepackage{bm}
\begin{document}
\draft
\renewcommand{\thefootnote}{\arabic{footnote}}

\title{Perturbative Unitarity Constraints on a Supersymmetric Higgs Portal} 
\bigskip
\author{Kassahun Betre, Sonia El Hedri and Devin~G.~E.~Walker}
\address{SLAC National Accelerator Laboratory, 2575 Sand Hill Road, Menlo Park, CA 94025, U.S.A.}

\begin{abstract}
\noindent
We place perturbative unitarity constraints on both the dimensionful and dimensionless couplings in the Next-to-Minimal Supersymmetric Standard Model Higgs Sector.  These constraints, plus the requirement that the singlino and/or Higgsino constitutes at least part of the observed dark matter relic abundance, generate upper bounds on the Higgs and neutralino/chargino mass spectrum.  We obtain an upper bound of 12 TeV for the charginos and neutralinos and 20 TeV for the heavy Higgses outside defined fine-tuned regions.  By using the NMSSM as a template, we describe a method which replaces naturalness arguments with more rigorous perturbative unitarity arguments to get a better understanding of when new physics will appear.

\end{abstract}

\maketitle
One of the most important questions in particle physics is to determine when new physics will appear.  Perturbative unitarity arguments have reliably answered this question throughout the development of the Standard Model (SM).  In particular, these arguments provided an upper bound on the SM Higgs mass~\cite{Dicus:1992vj,Lee:1977eg}.  This was possible because the Higgs quartic coupling was the only unknown parameter in the SM.  Thus, raising the SM Higgs mass increases this coupling until unitarity is violated.  %
For most models of new physics beyond the Standard Model (BSM), there are many new particles and couplings.  Thus, traditional 
unitarity arguments are insufficient and a definitive upper bound cannot be set.  In this letter, we describe how low-energy observables, such as the dark matter (DM) relic abundance, can work in concert with perturbative unitarity arguments to place upper bounds on particles predicted by BSM theories.  By doing so, we outline a method which replaces naturalness with perturbative unitarity arguments in order to determine a scale where new physics will appear~\cite{Walker:2013hka}.

We apply our arguments to the Higgs sector of the Next-to-Minimal Supersymmetric Standard Model (NMSSM)~\cite{Fayet:1974pd,Drees:1988fc}.  The NMSSM provides a~compelling framework for BSM physics.  The model naturally explains the observed SM Higgs mass~\cite{Aad:2012tfa,Chatrchyan:2012ufa}, provides a viable dark matter candidate, features gauge coupling unification and solves the hierarchy problem~\cite{Ellwanger:2009dp,Maniatis:2009re}.  To date, none of the new particles predicted by the NMSSM have been found.  In the following, %
we briefly review the NMSSM Higgs sector. We then describe how perturbative unitarity arguments can constrain both the dimensionless and dimensionful couplings in this sector.  
Next, we discuss the interplay between unitarity and DM relic density constraints and show the associated bounds.  Throughout, we discuss how the various constraints complement each other to 
directly constrain all of the free parameters in the NMSSM Higgs sector.  In~\cite{us}, we provide the full 
analysis including details on how the bounds vary when the bino and winos are light and mix with the singlino and Higgsinos.
\newline
\newline
\textbf{\underline{NMSSM Higgs Sector}:}  %
The scale invariant NMSSM Higgs superpotential is
\begin{equation}
    W = -\lambda\, \hat{S}\,(\hat{H}_u \cdot \hat{H}_d) + {\kappa \over 3}\, \hat{S}^3,
\label{eq:superpotential}
\end{equation} 
which adds two new dimensionless couplings, $\lambda$ and $\kappa$ to the MSSM. $\hat{H}_{u,d}$ are the up-~and down-type MSSM Higgs superfields  respectively.   $\hat{S}$ is the singlet superfield.  In this letter, we require all  the squarks, sleptons and gauginos to be heavy.  The soft-SUSY breaking scalar potential,
\begin{eqnarray}
V_\mathrm{soft} &=& m^2_{H_u}\, H^\dagger_u H_u + m^2_{H_d}\, H^\dagger_d H_d + m_S^2 \,S^\dagger S \label{eq:softsusy} \\
&-& \left( \lambda\, A_\lambda\, S \, H_u \cdot H_d  - {\kappa \over 3}\,A_\kappa \,S^3 + \mathrm{c.c.} \right), \nonumber
\end{eqnarray}
adds five additional dimensionful parameters. In total, the NMSSM Higgs sector contains two charged Higgses, 
three CP-even  neutral Higgses 
and three CP-odd neutral Higgses.  Also added are the fermionic superpartners, the singlinos and the Higgsinos.  
The lightest of the neutral Higgsinos/singlinos are stable and serve as dark matter candidates. 
%
The three neutral CP-even Higgses obtain vacuum expectation values (vevs).  Following the analysis of~\cite{Kanehata:2011ei}, we restrict to the parts of the moduli space that do not spontaneously break electric charge and are bounded from below.  The $\mu$ parameter is defined as $\mu = \lambda\, v_s$ where $v_s$ is the vev of the CP-even Higgs associated with the singlet superfield.  $\tan\beta$ is defined as $\tan\beta = v_u/v_d$ where $v_u$ and $v_d$ are the  vevs from the CP-even Higgses associated with the up- and down-type Higgs superfields.  
Requiring the correct electroweak vacuum leaves 
six free parameters,
\begin{equation}
\{\lambda, \,\kappa, \,\tan\beta, \,\mu, \,A_\lambda, \,A_\kappa \}.
\label{eq:parameters}
\end{equation}
%
We require the mass of the lightest neutral CP-even Higgs to equal the measured value of the SM Higgs mass~\cite{Aad:2012tfa,Chatrchyan:2012ufa}, %
%
\begin{equation}
m_h^2 = m_{H_1^0}^2  < m_Z^2\, \left( \cos^2 (2\beta)  + {2 \bigl| \lambda \bigr|^2 \sin^2 (2\beta) \over g_1^2 + g_2^2} \right).
\label{eq:SMhiggsconstraint}
\end{equation}
This tree-level constraint eliminates one degree of freedom from equation~\leqn{eq:parameters}.  Throughout this letter, our computations remain at tree-level.  Our results are not expected to vary significantly by including loop corrections to our unitarity or dark matter constraints.  Note however, finite and logarithmic radiative corrections from the decoupled sparticles can positively contribute to the squared SM Higgs mass.  Inherently, these contributions introduce additional SUSY breaking parameters associated with the sparticle sector.  This in turn forces $\tan\beta$ and $\lambda$ to be smaller and more tightly constrained in order for the lightest neutral CP-even Higgs mass to equal the measured SM Higgs mass.  %
%
%
To trigger the unitarity constraints, we consider the limit where the SUSY breaking scales are much larger\footnote{See equation~\ref{eq:scan} for the details of our parameter scan.} than the electroweak vev,
\begin{equation}
\mu,\,A_\lambda,\,A_\kappa \gg v.
\label{eq:decoupled}
\end{equation}
In this limit, the particles introduced by the NMSSM Higgs sector have masses that are simple functions of the parameters in equation~\leqn{eq:parameters}.    
\newline
\newline
\textbf{\underline{Perturbative Unitarity Arguments}:}  %
Unitarity is a basic pillar of quantum mechanics.  For unitary theories, the real and imaginary parts of a given scattering matrix are related through the optical theorem, which can %
%
lead to constraints on the parameters of a theory.  See e.g.,~\cite{Dicus:1992vj,Lee:1977eg} for a well-known example.  In this letter, we focus on %
$2 \to 2$ scattering processes involving only Higgs bosons.  We require  neutral CP-even initial and final states.  
Taking the scattering matrix $S = 1 + i\,T$, we project the $T$-matrix onto partial waves with total angular momentum $J$.  The strongest constraints on Higg-Higgs scattering processes come from the $J=0$ partial wave, and so, in this paper, we restrict to that case.  We label the projected $T$ matrix ${\cal T}^0_{if}$, where $i$ and $f$ label the initial and final state.  Because we assume no CP violating phases in the SUSY Higgs sector, $\mathcal{T}^0_{if}$ is fifteen dimensional~\cite{us}. 
To understand how unitarity can constrain the various parameters in the scattering matrix, consider 
the elastic scattering amplitude $h_s \,h_s  \to h_s \,h_s$ whose initial and final states are one element of the full scattering matrix.  Here $h_s$ is the Higgs boson of the singlet superfield $\hat{S}$.  The amplitude can be approximated as
\begin{equation}
\mathcal{\hat{T}} \sim \kappa^2 +  \kappa^2 A_\kappa^2\,\biggl( {1 \over s - m_{h_s}^2} +  {1 \over t - m_{h_s}^2} +  {1 \over u - m_{h_s}^2} \biggr),
\end{equation}
where $s$, $t$ and $u$ are the Mandelstam variables.  The first term on the RHS is the contribution to the amplitude from the $h_s$ quartic coupling.  
The other terms feature the tri-linear $h_s$ coupling, which is proportional to 
the dimensionful coupling $A_\kappa$~\cite{Ellwanger:2009dp,Maniatis:2009re}.  
In the limit where $s \to \infty$, only the quartic coupling survives.  Thus in this limit, the full scattering matrix can be used to directly constrain the dimensionless four-point couplings in the Higgs potential, $\lambda$ and $\kappa$.  
If $s$ is finite and equal to, e.g.~$s = 5 \,m_{h_s}^2$, a large ratio $A_\kappa^2/4\,m_{h_s}^2$ can also affect the size of $\mathcal{\hat{T}}$.  As we describe below, we always choose values for $s$ far enough from poles such that finite width effects are irrelevant.  Constraining the ratio of dimensionful parameters in~\leqn{eq:parameters} leaves one unconstrained dimensionful parameter.
%
\begin{figure}
    \centering
    \includegraphics[width=0.95\linewidth]{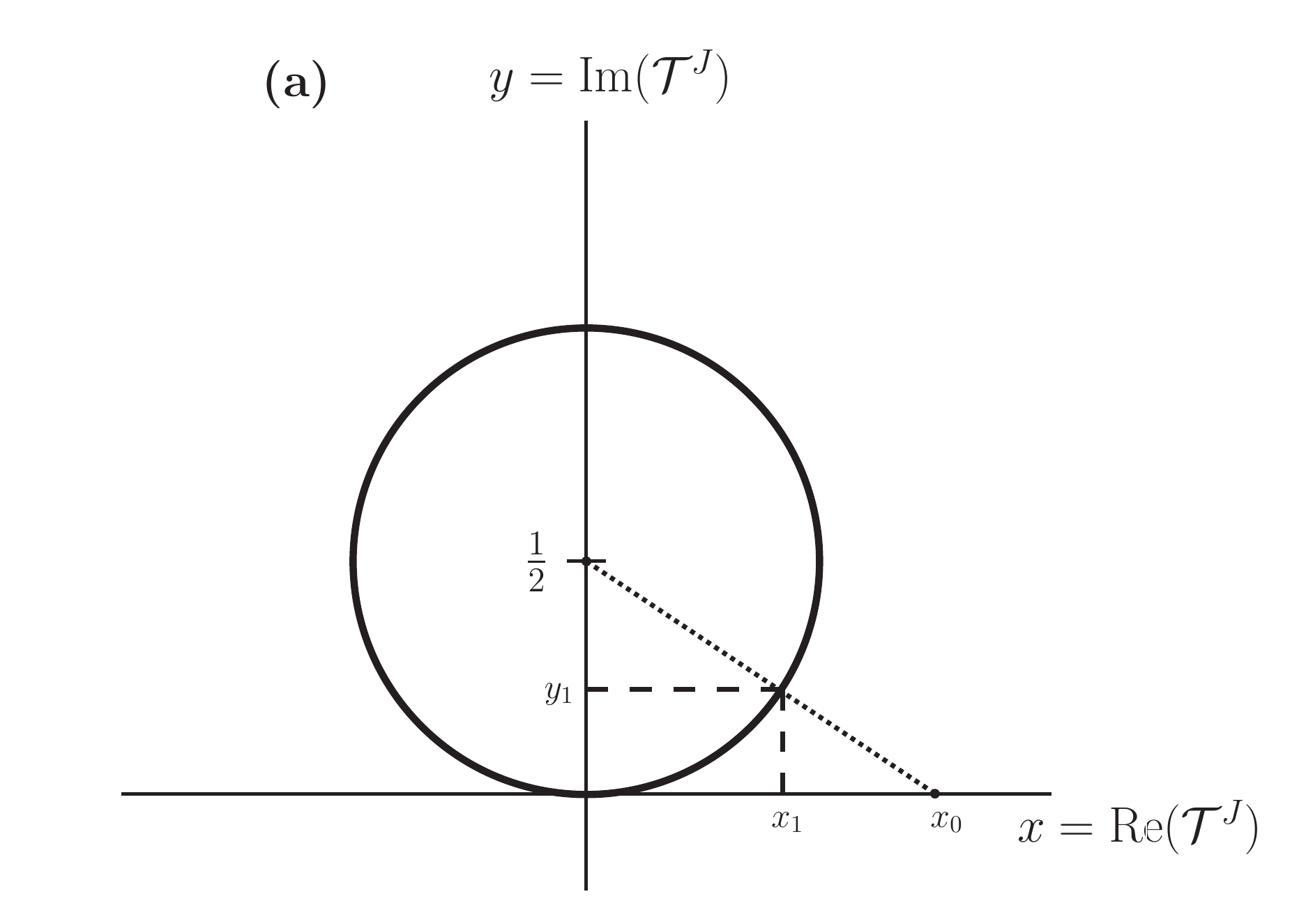} 
    \includegraphics[width=0.95\linewidth]{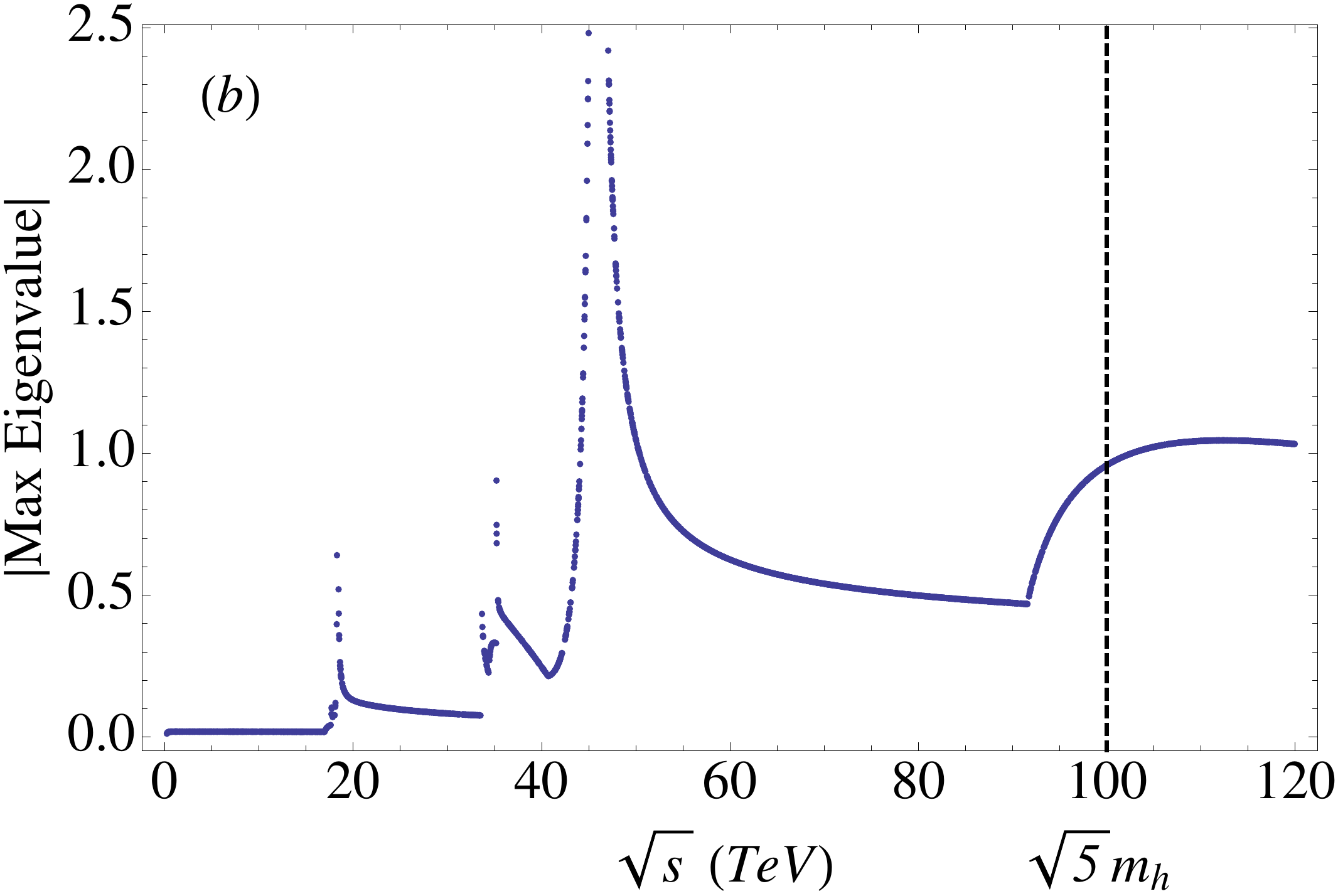} 
    \caption{\label{Fig: sqrtsscan}    (a) Argand diagram illustrating Schuessler and Zeppenfeld's arguments:  $\mathcal{T}^J_\mathrm{exact}$ is defined as the intersection of the Argand circle and a line connecting the center of the circle with the $J$th  partial wave  at tree-level, $x_0$.  $x_1$ and $y_1$ are the real and imaginary components of $\mathcal{T}^J_\mathrm{exact}$.  
    (b)  Absolute value of the largest eigenvalue as a function of $\sqrt{s}$ for an example point in the parameter space.  $m_H$ is the largest scalar mass; and the choice $s = 5\,m_H^2$ conservatively stays away from the unphysical poles.}
\end{figure}

To implement the unitarity bounds, we follow the approach of Schuessler and Zeppenfeld (SZ)~\cite{Schuessler:2007av,Schuessler:thesis}.  We diagonalize, $\mathcal{T}^0_{ij}$ which at tree-level yields real eigenvalues.  Thus in Figure 1a, our eigenvalues lie along the x-axis of the Argand diagram.  Of course, if we computed the scattering matrix to all orders of perturbation theory, the eigenvalues would lie on the Argand circle.  Each higher order correction moves our tree-level approximation towards the Argand circle, although circuitous routes are common~\cite{Aydemir:2012nz}.  
The essential part of SZ's arguments is to estimate the minimal correction needed to get from the x-axis to the Argand diagram regardless of the Mandelstam variable, $s$.  
Thus, given a tree-level partial wave expansion, they estimate the minimum amount of higher-order corrections needed over tree-level.  This is done by drawing a line from the center of the Argand circle to the tree-level partial wave expansion.  See Figure 1a.  The exact value of the scattering matrix to all orders of perturbation theory is taken to be the intersection between the line and the Argand circle.  Thus the minimal distance from the circle is 
\begin{equation}
    a ={ \bigl|\mathcal{T}_\mathrm{tree-level} - \mathcal{T}_\mathrm{exact}\bigr| \over \bigl|\mathcal{T}_\mathrm{tree-level}\bigr|}
\end{equation} 
which is normalized by the tree-level partial wave expansion.  %
We place unitarity bounds at the point where the tree-level eigenvalues must have a correction as large as $a \sim 41\%$.  
This correction is equivalent to $|\mathrm{Re}\, \mathcal{T}^0_{ii}| \leq 1/2$.  

An important complication to this story is choosing a suitable value of $\sqrt{s}$ that can maximize the Higgs-Higgs scattering matrix yet stay away from unphysical poles.  By looking at the eigenvalues of the Higgs-Higgs scattering matrices at various points in the parameter space (\ref{eq:parameters}), we conservatively choose $s = 5 m_H^2$ where $m_H$ is the largest scalar mass.  In Figure~1b, we plot the largest eigenvalue of the Higgs-Higgs scattering matrix for one point in parameter space to illustrate this choice.  When $s \to \infty$, our analysis is the same as the one used in~\cite{Lee:1977eg}.  We find the perturbative unitarity constraints force $| \lambda | \leq 3$ and $|\kappa | \leq 3$.  Because of all of the constraints listed above, 
only one unknown parameter in equation~\leqn{eq:parameters} remains unconstrained.
\newline
\newline
\textbf{\underline{Dark Matter Constraints}:}  %
In this letter, the singlino and/or Higgsinos are the thermal dark matter candidates.  In our extended analysis~\cite{us}, the winos and binos are included as potential dark matter candidates.  Working in the decoupling limit~\leqn{eq:decoupled}, the Higgsino and singlino masses are 
\begin{align}
    m_{\tilde{h}} = \mu + \mathcal{O}(v) &\,\,\,\,& m_{\tilde{s}} = 2\,\kappa\, \mu/\lambda + \mathcal{O}(v),
\end{align}
respectively.  If we raise $\mu$, the dark matter annihilation cross section must correspondingly increase in order to obtain a relic abundance equal to (or below) the measured value~\cite{Ade:2013zuv}.  A larger cross section implies larger couplings.  Eventually, for a large enough dark matter mass, the couplings that facilitate dark matter annihilation will be sufficiently large that they violate the perturbative unitarity constraints.  
Thus, because of the relic abundance, Higgs mass and unitarity requirements, all of the parameters in equation~\leqn{eq:parameters} are tightly constrained.  Notably, if the measured DM relic abundance is composed of a thermal and non-thermal component, then the thermal neutralino dark matter must annihilate more efficiently to get the right relic abundance.  In this case, the couplings which facilitate dark matter annihilations will have to be larger and violate perturbative unitarity constraints.  Thus, this scenario is more tightly constrained and the upper bound on the dark matter mass can be much smaller than the results presented in the next Section.%
%

We use MicrOmegas~\cite{Belanger:2013oya} to compute the relic abundance and direct detection constraints.  We require the relic abundance to be less than or equal to the central value measured by Planck~\cite{Ade:2013zuv} plus the three sigma error,
\begin{equation}
h^2 \Omega_c \leq 0.128.
\end{equation}
The exchange of light particles, such as the SM bosons, can lead to a large, non-perturbative enhancement of the dark matter annihilation cross section.  In~\cite{us}, we include 
these Sommerfeld enhancements for the wino/bino annihilation cross section.  Note, because the Higgsinos are coupled to light SM bosons, their annihilation cross section is potentially enhanced.  However, it should be noted that, e.g., one-loop corrections from stops and sbottoms induce Higgsino-Higgsino mass splittings that are often much larger than a GeV~\cite{Drees:1997uq,Lahanas:1993ib}.  These loop corrections eliminate the Sommerfeld enhancement for Higgsinos except in an extremely fine-tuned regime of parameter space where the effects of the loop corrections are cancelled.  Thus, in this letter we do not include Higgsino Sommerfeld enhancement in our results.  
%
Our results do include the Xenon1T sensitivity projections from~\cite{Aprile:talk} to estimate how the next generation of direct detection experiments will impact the available parameter space.
\newline
\begin{figure}
    \centering
    \includegraphics[width=0.95\linewidth]{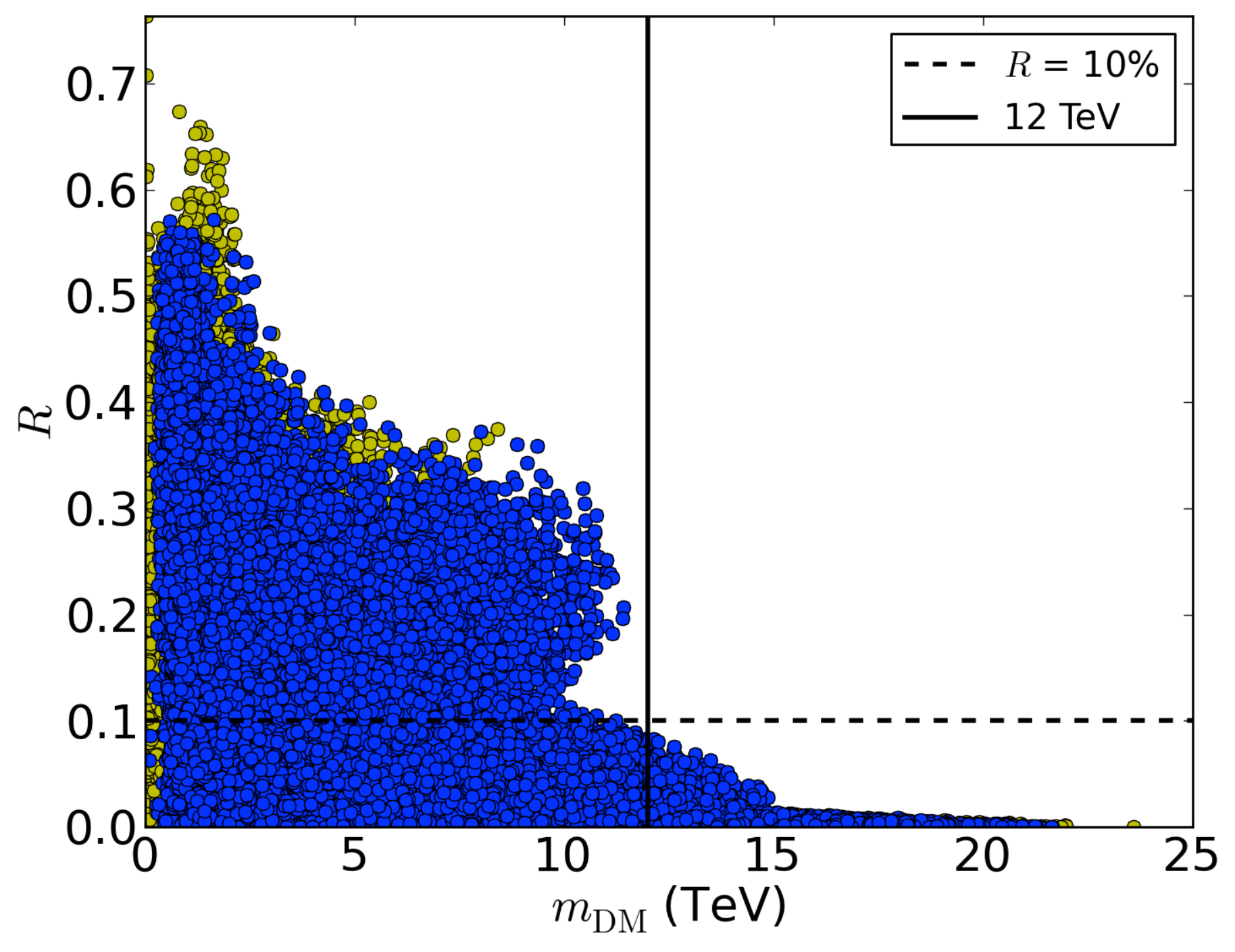} 
     \includegraphics[width=0.95\linewidth]{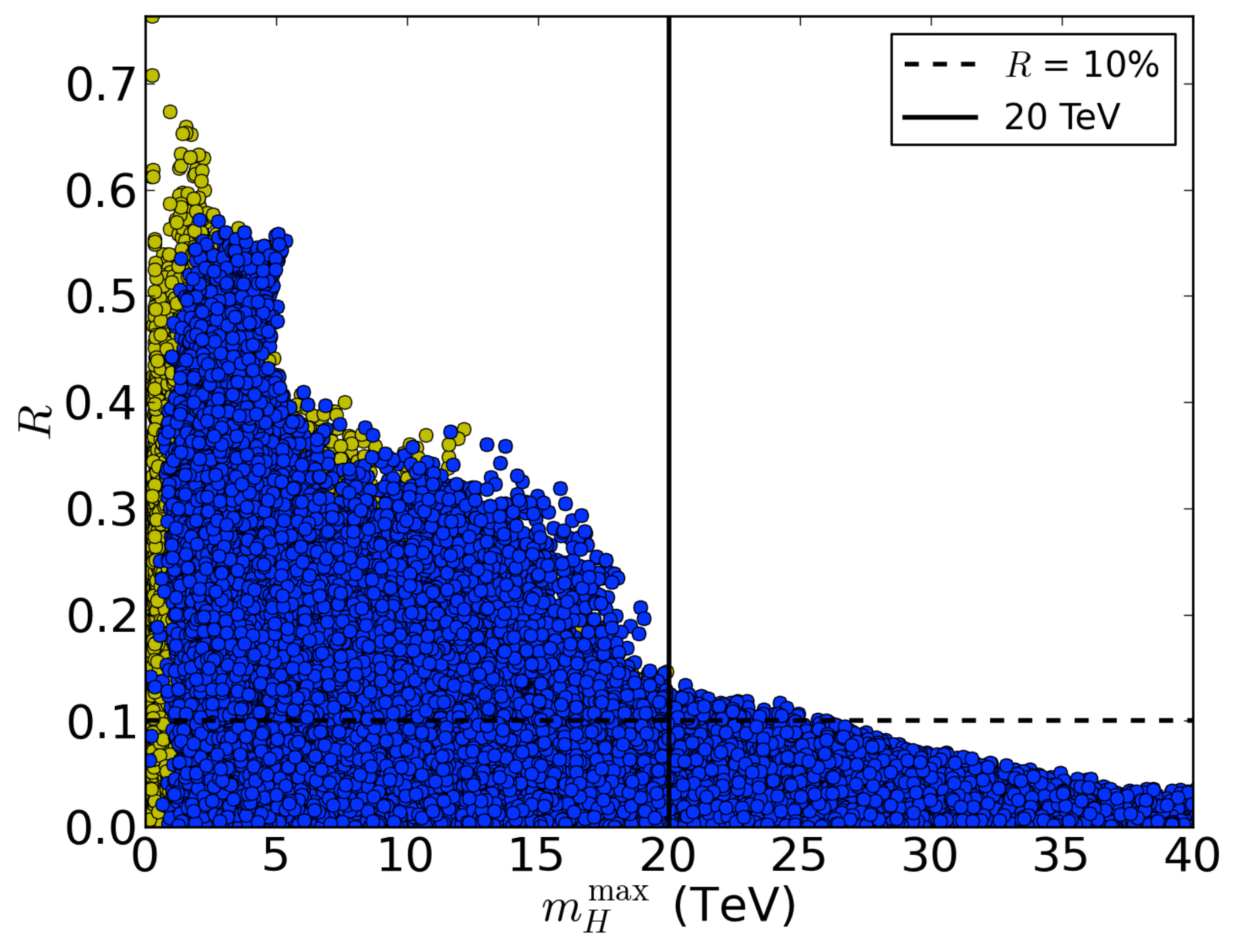}
     \caption{\label{Fig: sqrtsscan}   Points in parameter space which pass all constraints. $R$ is the fine-tuning parameter, defined in equation~\leqn{eq:fine-tuning}.  The dark yellow points are the Xenon 1T projections. (color online) The dashed horizontal line denotes 10\% fine-tuning of $R$.  (a) Higgsino and/or singlino dark matter constraints.  The ``chimney"-like relative maximum at 2 TeV is due to the saturating of the Higgsino dark matter.  The ``bump" between $5$ and $10$ TeV is composed of points that are dominantly singlino dark matter.  In (b), we plot heaviest CP-even, neutral Higgs constraints.}
\end{figure}
\begin{figure}
    \centering
    \includegraphics[width=0.95\linewidth]{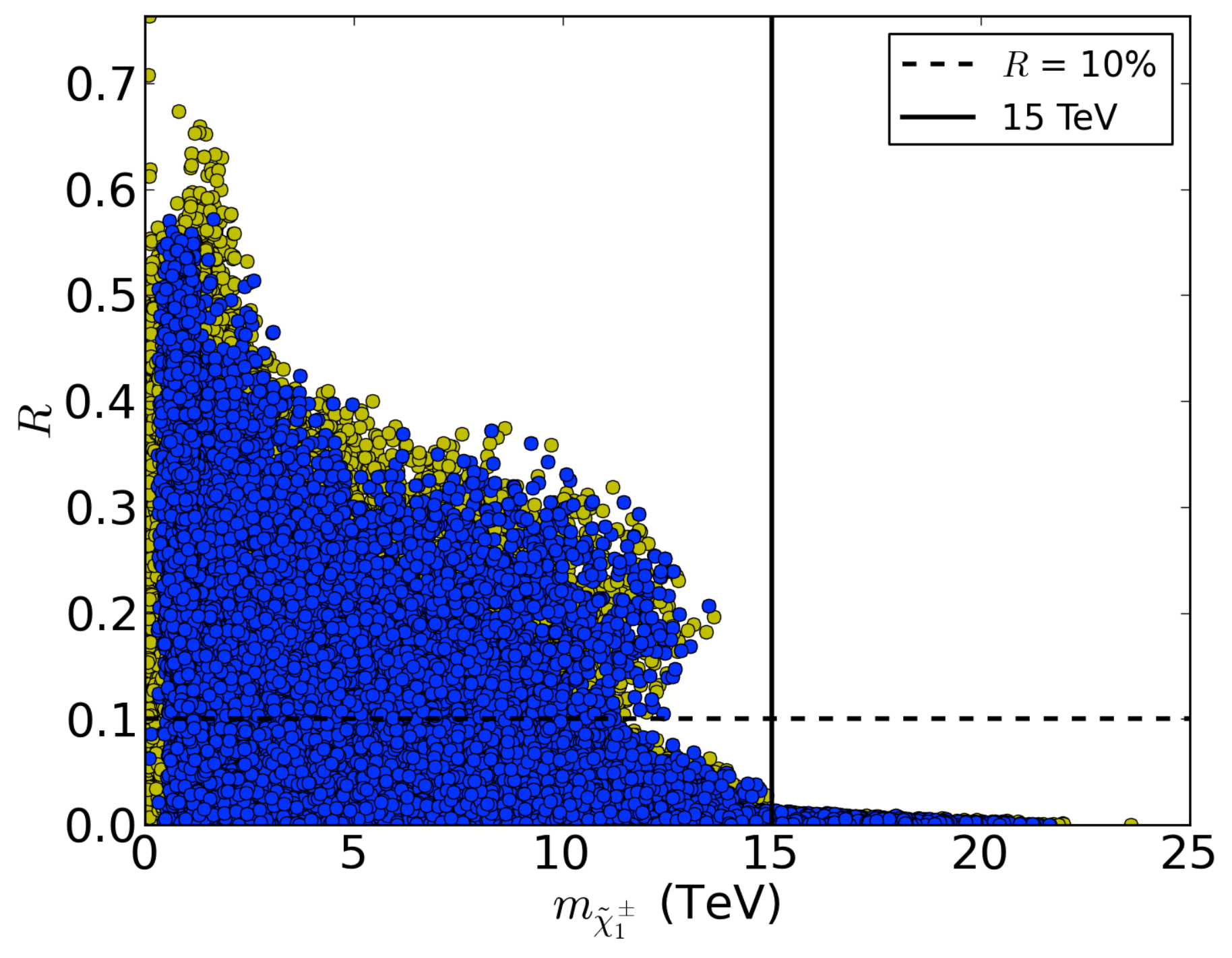} 
     \includegraphics[width=0.95\linewidth]{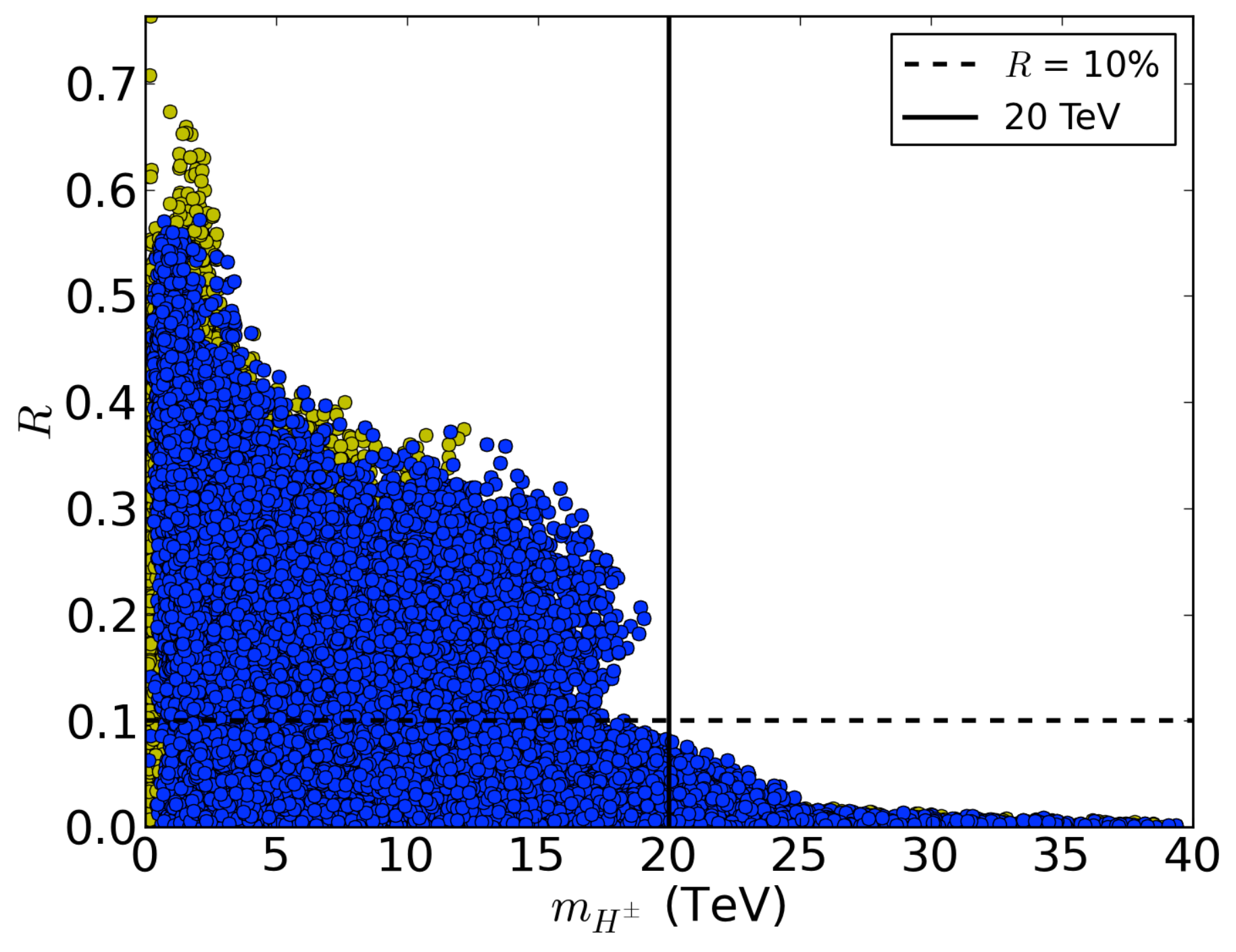}
     \caption{\label{Fig: sqrtsscan}  Points in parameter space which pass all constraints.$R$ is the fine-tuning parameter, defined in equation~\leqn{eq:fine-tuning}.  The dark yellow points are  the Xenon 1T projections. (color online)  The dashed horizontal line denotes 10\% fine-tuning of $R$.  In Figures (a) and (b) we plot the bounds on the heaviest chargino and heaviest charged Higgs.}
\end{figure}
\newline
\textbf{\underline{Results}:}  %
To determine the bounds on the NMSSM Higgs sector, we scan over the parameters in equation~\leqn{eq:parameters} with a uniform prior
\begin{align}
|A_{\lambda,\kappa}|, |\mu| \leq 40\,\,\mathrm{TeV} && |\lambda|, |\kappa| \leq 4 && 0 < \beta < {\pi \over 2}. \label{eq:scan}
\end{align}
Approximately $10^9$ points in parameter space were initially sampled.  The vacuum, SM Higgs Mass, relic abundance and perturbative unitarity constraints narrow that number to about $10^4$ points.  Some of the results are featured in Figs. 2 and 3.  

We define 
\begin{equation}
R = \mathrm{min}_i\biggl({|2\,m_\chi - m_{H_i} | \over m_{H_i}}  \biggr) \label{eq:fine-tuning}
\end{equation}
where $m_\chi$ is the neutralino dark matter mass and $m_{H_i}$ is the mass of the $i$th Higgs boson.  $i$ runs over all of the Higgs bosons in the theory.  $R$ effectively measures how close a given point in parameter space gets to the Higgs funnel region where the neutralinos resonantly annihilate.  In this region of parameter space, the denominator of the annihilation cross section can be extremely small therefore obviating the need for large couplings to get a large annihilation cross section.  Even in this ``fine-tuned" regime, the propagator denominator is regulated by the one-loop corrections.  Thus, a definitive upper bound exists~\cite{tom}.  This bound is reflected in the tails in our plots.  
Finally, our results in this letter do not consider Landau poles which can potentially provide a lower bound than the ones shown in Figures 2 and 3.  These are considered in~\cite{us}.
\newline
\newline
\textbf{\underline{Conclusions}}:  %
In this letter, we have argued that perturbative unitarity constraints can be augmented with constraints from low energy observables to set bounds on models of BSM physics.  In particular, we focused on the NMSSM Higgs sector and used the dark matter relic abundance to set a generic upper bound of $12$ TeV for the Higgsino/singlino Dark Matter and $20$ TeV for the heavy Higgs masses.  For this result we have required additional perturbative corrections to the tree-level scattering matrix to be no larger than $40$\%.  If we repeat the analysis described in this letter and require the perturbative corrections to be no larger than $20$\%, then the bound on all of the heavy Higgses dark matter is about $7$ TeV.  This is detailed in~\cite{us}.%
%

Naturalness arguments have dominated particle physicist's expectations of where new physics will appear.  However, since the new physics expected from these arguments has not been observed, the problem of finding the next scale of new physics remains.  With this letter we posit that novel unitarity arguments can be designed to determine new fundamental scales of physics.%
%
%

\vskip 0.2cm
\noindent
{\it Acknowledgments:}
We thank J.~Donoghue, A.~Nelson, M.~Peskin, T.~Rizzo and D.~Zeppenfeld for useful discussions.  We thank C.~Berger for accurate translations of~\cite{Schuessler:thesis} and M.~Peskin and T.~Rizzo reading drafts of this letter. This work is supported by the US Department of Energy, contract DE-ACO2-76SF00515. DW is supported in part by a grant from the Ford Foundation via the National Academies of the Sciences as well as the National Science Foundation under Grants No. NSF PHY11-25915 and NSF-PHY-0705682.  SEH is supported by a Stanford Graduate Fellowship.

\end{document}